\documentclass[aps,showpacs,twocolumn, prb]{revtex4-1}
\usepackage{hyperref}
\usepackage{graphicx}
\usepackage{amsmath,amssymb,amsfonts}
\usepackage{upgreek}
\usepackage{booktabs}
\usepackage{indentfirst}

\begin{document}

\title{The Chiral Non-Centrosymmetric Superconductors TaRh$_2$B$_2$ and NbRh$_2$B$_2$ }

\author{Elizabeth M. Carnicom$^{1*}$}
\author{Weiwei Xie$^2$}
\author{Tomasz Klimczuk$^3$}
\author{Jingjing Lin$^4$}
\author{Karolina G\'ornicka$^3$}
\author{Zuzanna Sobczak$^3$}
\author{Nai Phuan Ong$^4$}
\author{Robert J. Cava$^{1*}$}

\affiliation{$^1$Department of Chemistry, Princeton University, Princeton, New Jersey 08544, USA}
\affiliation{$^2$Department of Chemistry, Louisiana State University, Baton Rouge, Louisiana 70803, USA}
\affiliation{$^3$Department of Physics, Gdansk University of Technology, Gdansk 80-233, Poland}
\affiliation{$^4$Department of Physics, Princeton University, Princeton, New Jersey 08544, USA}

\begin{abstract}

The symmetry of a material's crystal structure has a significant effect on the energy states of its electrons. Inversion symmetry, for example, results in energetically degenerate electron energy bands for electrons with wave vectors \textbf{\textit{k}} and -\textbf{\textit{k}}, but, when spatial inversion symmetry is absent, this equivalence is no longer possible. In a non-centrosymmetric superconductor, such inequivalence has an important effect: the standard superconducting state, where electrons with opposite momenta form pairs on the Fermi surface, is not possible. A handful of such materials is known; they display different degrees of influence of this lack of inversion symmetry on their superconducting properties. The effect of crystal structure chirality on the properties and applications of superconductors, on the other hand, is little discussed. Here we report the new isostructural non-centrosymmetric superconductors TaRh$_2$B$_2$ and NbRh$_2$B$_2$, which have a previously unreported crystal structure type. Not only do these materials lack inversion symmetry, but their crystal structure is also chiral; in other words, they can exist in right-handed or left-handed forms. Unlike most superconductors, their upper critical magnetic fields extrapolated to 0 K exceed the Pauli limit, which is often taken as the first indication that a superconducting material is anomalous. We propose that these materials represent a new kind of platform on which the effects of handedness on superconductors and their devices can be tested.  

\end{abstract}
\maketitle

\section{Introduction}

The discovery of superconductivity in CePt$_3$Si\cite{CePt3Si_1, CePt3Si_2} has sparked interest in superconductors with non-centrosymmetric crystal structures\cite{NCS_Rev1} and their anomalous character.\cite{CeIrSi3,UIr}  By far, most superconductors reported to date possess inversion symmetry, believed to be a favorable trait in a superconducting material. When a superconductor has an inversion center, it can be classified as having either spin singlet or spin triplet pairing since the resulting spin degeneracy is protected by the inversion.\cite{tripletSC_2,tripletSC_3}  However, for non-centrosymmetric crystal structures, conventional Cooper pairs can no longer form since a state on the Fermi surface with momentum \textbf{\textit{k}} does not have a degenerate pair at -\textbf{\textit{k}}.\cite{HelicalVortex} When inversion symmetry is absent, anti-symmetric spin-orbit coupling (ASOC) breaks the degeneracy, and states of mixed parity become possible, which can result in complicated spin structures.\cite{ElectronCorr,ThCoC2}  Such an admixture of spin states results in quasiparticle band structures that are topologically nontrivial, which in turn results in a material with protected zero-energy states at the surface or edges, similar to topological insulators.\cite{topological3, topological2}

The violation of parity conservation can lead to various other exotic behaviors, like magnetoelectric coupling\cite{polarSC} or anomalous upper critical field ($\mu_0$\textit{H}$_\text{c2}$) values.\cite{NCS_Rev1} Some non-centrosymmetric superconductors have been shown to have upper critical fields that exceed the Pauli limit ($\mu_0$\textit{H}$^\text{Pauli}$ = 1.85*\textit{T}$_\text{c}$)\cite{PM_limitHc2,CeRhSi3}, a limit that is based on the maximum magnetic field that will not break apart a singlet Cooper pair at low temperatures.\cite{Pauli_Limit6} Often  the interplay between breaking inversion symmetry and superconductivity becomes difficult to unravel when other factors like heavy fermion behavior\cite{CeCoGe3}, as seen in CePt$_3$Si\cite{CePt3Si_1}, or magnetism\cite{UIr} are present. Studying superconductivity in transition-metal containing non-centrosymmetric compounds such as Mg$_{10}$Ir$_{19}$B$_{16}$,\cite{Mg10Ir19B16,Mg10Ir19B16_2} Nb$_{0.18}$Re$_{0.82}$,\cite{NbRe_paper1} \textit{T}$_2$Ga$_9$ (\textit{T} = Rh, Ir),\cite{Rh2Ga9} Li$_2$\textit{T}$_3$B (\textit{T} = Pd, Pt),\cite{LiPdB_1, LiPdB_2} or other non-magnetic analogs,\cite{BaPtSi3, SrAuSi3} such as LaPt$_3$Si\cite{LaPt3Si}, is important because factors like the strong correlation of  \textit{f}-electrons and magnetic interactions are absent. Although Mg$_{10}$Ir$_{19}$B$_{16}$,\cite{Mg10Ir19B16,Mg10Ir19B16_2} CePt$_3$Si,\cite{CePt3Si_1, CePt3Si_2} and Nb$_{0.18}$Re$_{0.82}$\cite{NbRe_paper1} are non-centrosymmetric, their crystal structures are non-chiral, in other words they are non-enantiomorphic. Li$_2$\textit{T}$_3$B (\textit{T} = Pd, Pt)\cite{LiPdB_1, LiPdB_2} and Mo$_3$Al$_2$C\cite{Mo3Al2C} are both non-centrosymmetric and have chiral crystal structures, but the potential effects of their chirality on superconductivity was not noted. (Chiral crystal structures must also be non-centrosymmetric, but non-centrosymmetric crystal structures are not necessarily chiral.) Interestingly, Sr$_2$RuO$_4$\cite{Sr2RuO4_cryst,ST_Sr2RuO4} and UPt$_3$\cite{UPt3}, which are centrosymmetric and therefore achiral, are, nonetheless, two of the main candidates believed to display chiral superconductivity.\cite{chiralSC} Li$_2$Pt$_3$B has also been proposed\cite{LiPdB_1,Li2Pt3B_ST} as a candidate for chiral superconductivity. In order for a superconductor to be chiral, the phase of the superconducting gap function, $\Delta$(\textbf{\textit{k}}), must wind in either a counter-clockwise or a clockwise direction as \textbf{\textit{k}} moves along the Fermi surface.\cite{chiralSC}

In this study, we report our discovery and investigation of the symmetrically chiral, non-centrosymmetric superconductors TaRh$_2$B$_2$ and NbRh$_2$B$_2$; reporting their characterization through single-crystal X-ray diffraction, temperature-dependent electrical resistivity, magnetic susceptibility, and specific-heat measurements. They are isostructural, both forming in the same chiral non-centrosymmetric space group \textit{P} 3$_1$ (No. 144). We show that some of their properties are consistent with normal conventional superconducting behavior, but we also find that both TaRh$_2$B$_2$ and NbRh$_2$B$_2$ display anomalous upper critical fields (\textit{H}$_\text{c2}$(0)) values that exceed the Pauli limit. We therefore propose that these chiral, non-centrosymmetric materials may display exotic pairing symmetries in the superconducting state. 
	 
\section{Experimental Methods}
 
\textbf{Experimental Design}
The starting materials for the synthesis of TaRh$_2$B$_2$ and NbRh$_2$B$_2$ bulk materials were elemental tantalum ($>$99.9$\%$, 100 mesh, Alfa), niobium ($>$99.9$\%$, 200 mesh, Aldrich), rhodium ($>$99.9$\%$, 325 mesh, Alfa), and boron (submicron particles, Callery Chem.). Powders of the starting materials Ta/Nb, Rh, and B were weighed out in a 1:1.9:2.1 ratio, ground using a mortar and pestle, and pressed into a pellet. The samples were then wrapped in Ta foil, placed in an alumina crucible, and heated in a high vacuum furnace to 1200 $^{\circ}$C for 6 hr. Variation from the above loading compositions or heating temperature led to the presence of secondary phases in larger amounts. Attempts to arc-melt TaRh$_2$B$_2$ followed by annealing at 1100 $^{\circ}$C for 1 week in an evacuated silica tube did not result in a single-phase sample. Samples of TaRh$_2$B$_2$ and NbRh$_2$B$_2$ are stable in air and do not decompose over time. 

\textbf{Crystal Structure Characterization}
The purity of the samples studied was checked using room temperature powder X-ray diffraction (pXRD) using a Bruker D8 Advance Eco Cu K$_\alpha$ radiation ($\lambda$ = 1.5406 $\textsc{\AA}$) diffractometer with a LynxEye-XE detector. Single crystals from an arc-melted and annealed sample of TaRh$_2$B$_2$ were mounted on the tips of Kapton loops. A Bruker Apex II X-ray diffractometer with Mo K$_{\alpha1}$ radiation ($\lambda$ = 0.71073 $\textsc{\AA}$) was used to collect room temperature intensity data. The data were collected over a full sphere of reciprocal space with 0.5$^{\circ}$ scans in $\omega$, 10s per frame of exposure time, and a 2$\theta$ range from 4$^{\circ}$ to 75$^{\circ}$. The SMART software was used for acquiring all data and the SAINT program was used to both extract intensities and to correct for polarization and Lorentz effects. XPREP was used to perform numerical absorption corrections.\cite{SHELXTL} Twinning of the unit cell was tested. The crystal structure of TaRh$_2$B$_2$ was solved using direct methods and refined by full-matrix least-squares on F$^2$ with the SHELXTL package.\cite{SHELX} A total of 25 space groups were tested according to the observed Laue symmetry and the space group was determined to be \textit{P} 3$_1$. To the best of our knowledge, TaRh$_2$B$_2$ has a unique structure type. The crystal structure images were created in the program VESTA.\cite{VESTA} Since boron is such a light element, and therefore difficult to quantify using X-ray diffraction, the B content determined in the refinement was tested further through synthesis of materials of different boron contents. Such syntheses resulted in significant amounts of impurity phases. Finally, the surfaces of Ta-Rh-B samples were analyzed for their Ta and Rh content using an FEI Quanta 250FEG scanning electron microscope equipped with an Apollo-X SDD energy-dispersive spectrometer (EDS). EDAX TEAM$^\text{TM}$ software was used to process the EDS data by using standardless quantitative analysis. The EDS data confirmed a Ta:Rh ratio of 1:2, consistent with the results from the single crystal refinement. The FullProf Suite program was used to perform LeBail refinements on bulk samples of both TaRh$_2$B$_2$ and NbRh$_2$B$_2$ using Thompson-Cox-Hastings pseudo-Voigt peak shapes. Lattice parameters determined from single crystal data were consistent with those determined from the LeBail fit for TaRh$_2$B$_2$. The same structure with slightly different cell parameters was also found to index the diffraction pattern for NbRh$_2$B$_2$. 

\textbf{Superconducting Property Measurements}
A Quantum Design Physical Property Measurement System (PPMS) Dynacool was used to measure the temperature and field-dependent magnetization and temperature-dependent electrical resistivity of TaRh$_2$B$_2$ and NbRh$_2$B$_2$ using a vibrating sample magnetometer (VSM) and resistivity option. Both zero-field cooled (ZFC) and field-cooled (FC) magnetic susceptibility measurements were taken from 1.7 K - 10 K for the Ta-variant and from 1.7 K - 12 K for the Nb-variant with a 10 Oe applied magnetic field. The field-dependent magnetization was measured at various temperatures for both TaRh$_2$B$_2$ and NbRh$_2$B$_2$ for fields in the range 0 - 150 Oe for the Ta-variant and 0 - 350 Oe for the Nb-variant. The temperature-dependent electrical resistivity measurements were carried out using a standard four-probe method from 300 K - 1.7 K with an applied current of 2 mA and zero applied magnetic field. The resistivity was then measured in the low temperature region from 1.7 K - 7 K for the Ta-variant and 1.7 K - 9 K for the Nb-variant with applied magnetic fields varying from 0 T to 9 T in 0.5 T increments. A 14 T Quantum Design PPMS was used to measure the high field temperature dependent resistivity of both NbRh$_2$B$_2$ and TaRh$_2$B$_2$ with a 2 mA applied current in 0.5 T increments.  Specific heat data for both TaRh$_2$B$_2$ and NbRh$_2$B$_2$ were collected on a Quantum Design PPMS Evercool II with applied magnetic fields of 0 T and 9 T.

\textbf{Electronic Structure Calculations}
The band structures (BS) and density of states (DOS) of TaRh$_2$B$_2$ and NbRh$_2$B$_2$ were calculated using Wien2K with LDA-type pseudopotentials. The structural lattice parameters obtained from single crystal diffraction experiments for TaRh$_2$B$_2$ were used for the calculation. Spin-orbit coupling was included for all atoms. Reciprocal space integrations were completed over an 8x8x4 Monkhorst-Pack \textit{k}-points mesh. The convergence criterion for a self-consistent calculation was taken as 1.0x10$^{-4}$ eV. 

\section{Results and Discussion}
\begin{table}
\caption{Single crystal crystallographic data for TaRh$_2$B$_2$.}
\begin{tabular}{c c} 
\hline 
Chemical Formula & TaRh$_2$B$_2$  \\
\hline  
Temperature(K)				&	293(2)K							\\
F.W. (g/mol); 				& 408.39 							\\  
Space group; $Z$ 			& $P$ 3$_1$ (No. 144);3 		\\  
$a$ (${\textsc{\AA}}$)  	& 4.6980(7)							\\ 
$c$ (${\textsc{\AA}}$)   	& 8.770(2)						\\

$V$ (${\textsc{\AA}^3}$) 	& 167.63(6)						\\

\textit{hkl} ranges		 	& $-$7 $\le$ $hk$ $\le$ 7			\\
							& $-$13 $\le$ $l$ $\le$ 13			\\
Absorption Correction	 	& Numerical							\\
Extinction Coefficient		& 	0.0030(6)				\\
$\theta$ range (deg.)	 	& 5.011$-$33.159			\\
No. reflections; $R_{int}$	& 870; 0.0466						\\
No. independent reflections & 		760					\\
No. parameters				& 		48						\\
$R_1$; $wR_2$ (I$>$2$\delta$(I))&	0.0331; 0.0587		\\
$R_1$; $wR_2$ (all I)		&  0.0430; 0.0626				\\
Goodness of fit				& 	1.146						\\
Diffraction peak and hole (e$^-$/${\textsc{\AA}^3}$)	& 5.946; -5.241\\
\hline
\hline
\end{tabular}
\label{Table1} 
\end{table}

\textbf{Crystal Structure}
\begin{figure}[t]
\includegraphics[scale = 0.40]{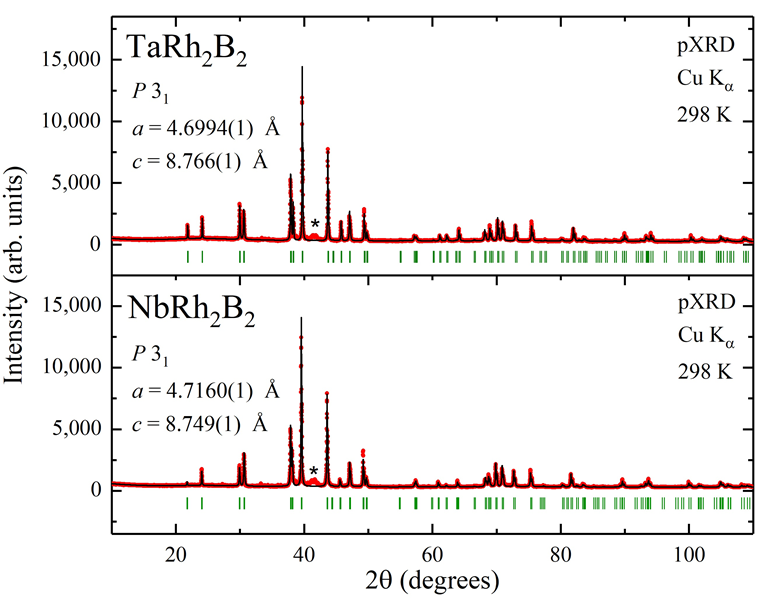}
\caption{Room temperature powder X-ray diffraction pattern showing a LeBail fit for the new superconducting phases TaRh$_2$B$_2$ (top) and NbRh$_2$B$_2$ (bottom). The experimentally observed data are shown with red circles, the calculated pattern is shown with a black line, and the green vertical marks indicate the expected Bragg reflections for space group \textit{P} 3$_1$ (No. 144). Impurity peaks are marked with asterisks.}
\label{Fig1}
\end{figure}

\begin{figure}[b]
\includegraphics[scale = 0.6]{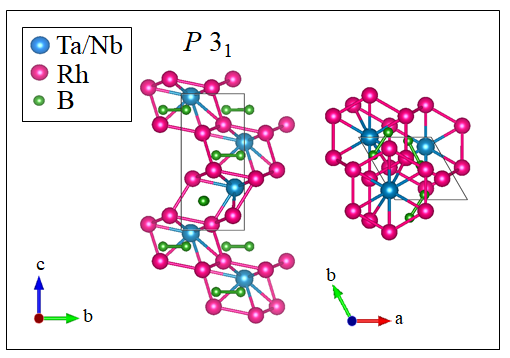}
\caption{The crystal structure of chiral and non-centrosymmetric TaRh$_2$B$_2$ and isostructural NbRh$_2$B$_2$ viewed along the \textit{a}-direction, emphasizing the 3$_1$ screw axis (left) and along the \textit{c}-direction (right), emphasizing the lack of inversion symmetry. Tantalum/niobium is shown in blue, rhodium is shown in pink, and boron is shown in green.}
\label{Fig2}
\end{figure}

	Single crystal diffraction was used to determine the crystal structure of the new superconductor TaRh$_2$B$_2$, which crystallizes in the chiral, non-centrosymmetric space group \textit{P} 3$_1$ (No. 144), where \textit{a} = 4.6980(7) $\textsc{\AA}$ and \textit{c} = 8.770(2) $\textsc{\AA}$.  The crystal structure is reported here for the first time. The space group and lattice parameters from the single crystal refinement were used as a starting point for the LeBail fits of the room temperature pXRD patterns in Fig. \ref{Fig1} for TaRh$_2$B$_2$ (top) and isostructural NbRh$_2$B$_2$ (bottom). Only standard solid-state synthesis methods resulted in making the nearly single-phase bulk samples whose powder patterns are shown in Fig.\ref{Fig1}. The 3$_1$ screw axis in the structure, which results in its chirality, can be seen in Fig.\ref{Fig2} (left) when the structure is viewed along the \textit{a}-direction. In addition, viewing the structure along the \textit{c}-direction (Fig.\ref{Fig2}, right) emphasizes the non-centrosymmetric nature of the material. A summary of the single crystal structure refinement for TaRh$_2$B$_2$ is given in Table \ref{Table1}. The atomic coordinates from the crystal structure refinement are shown in Table \ref{Table2}. The new superconducting materials TaRh$_2$B$_2$ and isostructural NbRh$_2$B$_2$ have crystal structures that are both chiral and non-centrosymmetric. 
	
\begin{table}
\caption{Atomic coordinates and equivalent isotropic displacement parameters of TaRh$_2$B$_2$ at 293(2) K. U$_\text{eq}$ is defined as one-third of the trace of the orthogonalized U$_\text{ij}$ tensor (${\textsc{\AA}^2}$).}
\begin{tabular}{ccccccc} 
\hline
\hline 
Atom & Wyck. & Occ. & $\textit{x}$ & $\textit{y}$ & $\textit{z}$ &U$_\text{eq}$\\ 
\hline 
Ta	& 3\textit{a} 		& 1 & 0.8481(2)				 & 0.8481(2)      	& 0.3131(2) &0.0041(2) \\
Rh1 	& 3\textit{a}  	&1 & 0.6638(4) 		& 0.4824(4) & 0.0447(3) & 0.0058(2)\\  
Rh2 	& 3\textit{a}  	& 1 & 0.3324(4)  	& 0.8190(4) & 0.1072(2) &0.0058(3)\\ 
B1		&3\textit{a}  	& 1 & 0.627(7)   & 0.550(8)	& 0.549(3) & 0.012(5)\\ 
B2 	& 3\textit{a}  		& 1 & 0.288(9)   &  0.347(8) &0.215(3) &0.028(5) \\ 

\hline
\hline 
\end{tabular}
\label{Table2} 
\end{table}

\begin{figure}[t]
\includegraphics[scale = 0.66]{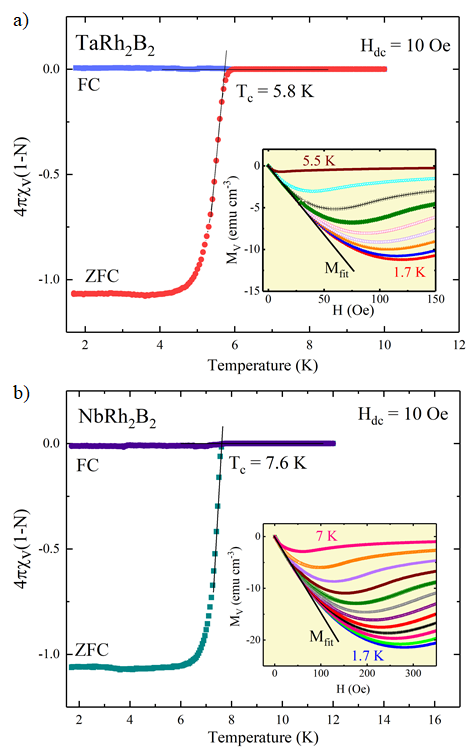}
\caption{Zero-field cooled and field cooled temperature-dependent magnetic susceptibility $\chi_V$(\textit{T}) for a) TaRh$_2$B$_2$ and b) NbRh$_2$B$_2$ measured in an applied magnetic field of 10 Oe. Inset: Field-dependent volume magnetization (\textit{M}$_V$) measured at various temperatures below \textit{T}$_\text{c}$. The fit lines \textit{M}$_\text{fit}$ were used to estimate the demagnetization factor, \textit{N}, for both materials. These values were used to correct the magnetic susceptibility data in the main panels.}
\label{Fig3}
\end{figure}

\textbf{Magnetic Susceptibility}
	To characterize the superconductors TaRh$_2$B$_2$ and NbRh$_2$B$_2$, the field-dependent volume magnetization (\textit{M}$_V$) was measured at 1.7 K (in the superconducting state) for both TaRh$_2$B$_2$ (Fig.\ref{Fig3}a, inset) and NbRh$_2$B$_2$ (Fig.\ref{Fig3}b, inset) and fitted to a line in the low field region (\textit{M}$_\text{fit}$ = \textit{bH} + \textit{a}). The equation $-b = \frac{1}{4\pi(1-N)}$ was used to estimate the value of the demagnetization factor, \textit{N} (which is based on the sample shape and orientation with respect to the applied magnetic field), using the slope of the fitted line, \textit{b}. Further characterization was carried out through both zero-field cooled and field-cooled temperature-dependent magnetic susceptibility measurements with a 10 Oe applied field. The measurement was taken from 1.7 K - 10 K for TaRh$_2$B$_2$ (Fig.\ref{Fig3}a, main panel) showing a superconducting critical temperature of 5.8 K. The \textit{T}$_\text{c}$ was determined as the intersection between the normal state of the magnetic susceptibility extrapolated to lower temperature with the line corresponding to the steepest slope of the diamagnetic signal (indicated by black solid lines).\cite{Tc_choice} ZFC and FC measurements were also performed on the Nb-variant (Fig.\ref{Fig3}b, main panel) from 1.7 K - 12 K, showing a \textit{T}$_\text{c}$ = 7.6 K. Both the ZFC and FC measurements were corrected for \textit{N} = 0.504 for the Ta-variant and \textit{N} = 0.448 for the Nb-variant. The resulting diamagnetic signal is only slightly less than the ideal value of 4$\pi\chi_V$(1-\textit{N}) = -1 for both TaRh$_2$B$_2$ and NbRh$_2$B$_2$. The observed critical temperatures in the susceptibility measurements are consistent with both the specific heat and resistivity measurements (discussed next) for each superconducting material. There is a very weak superconducting signal observed for the field-cooled measurement for each material, as expected due to the polycrystalline nature of both samples.

\begin{figure}[b]
\includegraphics[scale = 0.65]{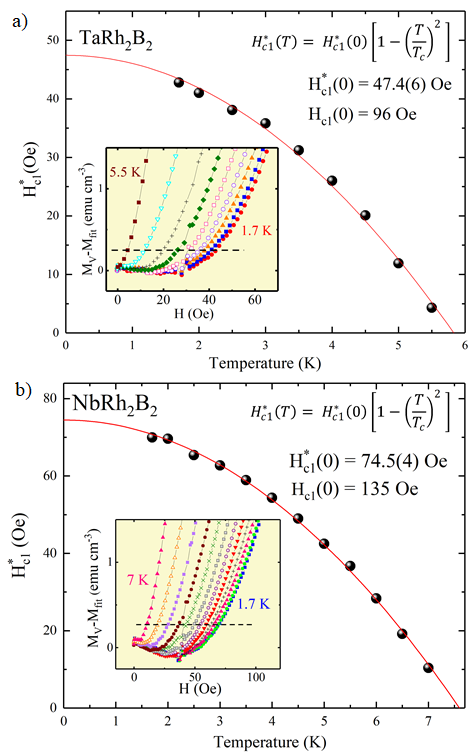}
\caption{Estimation of the lower critical field \textit{H}$_\text{c1}^{*}$ vs. temperature for a) TaRh$_2$B$_2$ and b) NbRh$_2$B$_2$. Inset: Plot of \textit{M}$_V$ - \textit{M}$_\text{fit}$ used to obtain data points to estimate \textit{H}$_\text{c1}^{*}$.}  
\label{Fig4}
\end{figure}
 
\textbf{Magnetization}
The new superconductors TaRh$_2$B$_2$ and NbRh$_2$B$_2$ were further studied with field-dependent magnetization (\textit{M}$_V$) measurements at different temperatures below the critical temperature as seen in Fig.\ref{Fig3}a (inset) for the Ta-variant and Fig.\ref{Fig3}b (inset) for the Nb-variant. The low-field linear fits to the magnetization data (\textit{M}$_\text{fit}$), discussed previously, were used to construct the \textit{M}$_V$-\textit{M}$_\text{fit}$ plot in Fig.\ref{Fig4}a (inset) for TaRh$_2$B$_2$ and NbRh$_2$B$_2$ (Fig.\ref{Fig4}b, inset). The field at which the magnetization begins to deviate from a linear response (indicated by the black dashed line in each inset) is the uncorrected lower critical field, \textit{H}$_\text{c1}^{*}$, for that temperature. All the \textit{H}$_\text{c1}^{*}$ values with the corresponding temperatures were plotted in Fig.\ref{Fig4}a (main panel) for TaRh$_2$B$_2$ and in Fig.\ref{Fig4}b (main panel) for NbRh$_2$B$_2$ and fitted to the following equation,
\begin{equation}
H_{c1}^{*} (T) = H_{c1}^{*}(0)\left[ 1-\left( \frac{T}{T_c}\right) \right] ^2,
\label{Eq1}
\end{equation}
where \textit{H}$_\text{c1}^{*}$(0) is the lower critical field at 0 K and \textit{T}$_\text{c}$ is the superconducting critical temperature. \textit{H}$_\text{c1}^{*}$(0) was calculated to be 47.4(6) Oe for TaRh$_2$B$_2$ and 74.5(4) Oe for NbRh$_2$B$_2$. After correcting for the demagnetization factor for each sample, \textit{H}$_\text{c1}$(0) was calculated to be 96 Oe for the Ta-variant and 135 Oe for the Nb-variant.

\begin{figure}[t]
\includegraphics[scale = 0.64]{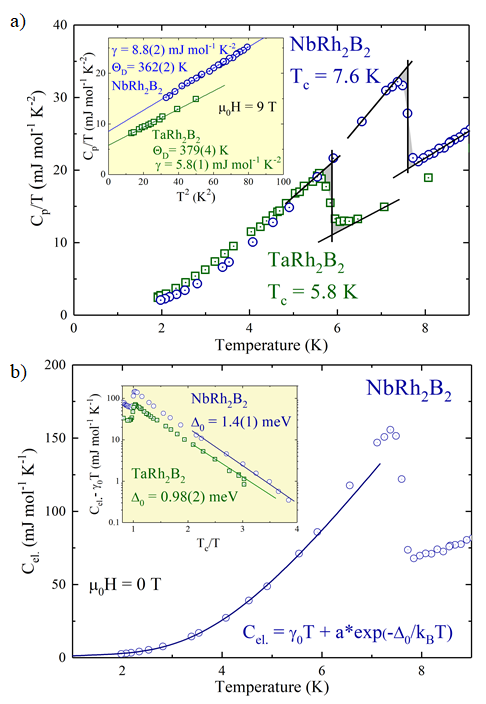}
\caption{a) \textit{C$_p$}/\textit{T} vs. \textit{T} plotted from 0 K - 9 K for TaRh$_2$B$_2$ (green open squares) and NbRh$_2$B$_2$ (blue open circles) measured in zero applied field where the solid black lines outline the equal area construction shown with grey shading. This construction is used for the estimation of \textit{T}$_\text{c}$ and the superconducting jump $\Delta$\textit{C}/\textit{T}$_\text{c}$. Inset: \textit{C$_p$/T} vs. \textit{T}$^2$ measured in a 9 T field (in the normal state) fitted to \textit{C$_p$}/\textit{T} = $\gamma$ + $\beta$\textit{T}$^2$. b) Temperature dependent electronic specific heat \textit{C}$_\text{el.}$ for NbRh$_2$B$_2$ below 9 K. The solid curve through the data is a fit by a one-gap BCS model for superconductivity. Inset: \textit{C}$_\text{el.}$-$\gamma_0$\textit{T} vs. normalized temperature (\textit{T}$_\text{c}$/\textit{T}) for NbRh$_2$B$_2$ and TaRh$_2$B$_2$.}
\label{Fig5}
\end{figure}

\textbf{Specific Heat and Observed Superconducting Parameters}
	The temperature-dependence of \textit{C$_p$/T} is plotted for both TaRh$_2$B$_2$ and NbRh$_2$B$_2$ in Fig.\ref{Fig5}a (main panel) under zero applied magnetic field measured from 2 K - 9 K, showing a large anomaly in the specific heat for each superconducting material. Using equal-entropy constructions of the idealized specific heat capacity jump (grey shading), the \textit{T}$_\text{c}$ was determined to be 5.8 K for the Ta-variant and 7.6 K for the Nb-variant, consistent with the magnetic susceptibility and resistivity data.
	 
\begin{figure*}[tbp]
\includegraphics[scale = 0.5]{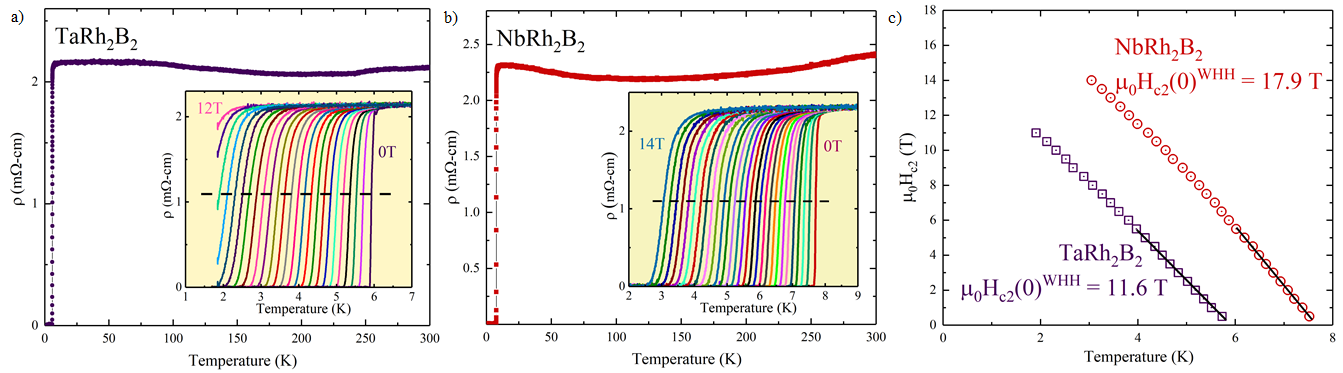}
\caption{Temperature-dependent electrical resistivity of a) TaRh$_2$B$_2$ and b) NbRh$_2$B$_2$ measured from 300 K - 1.7 K with zero applied magnetic field. Inset: Plot of the dependence of the superconducting transition on applied magnetic field measured from a) 1.7 K - 7 K for the Ta-variant in applied fields from 0 T - 12 T and b) from 1.7 K - 9 K for the Nb-variant in applied magnetic fields from 0 T to 14 T in 0.5 T increments. The dashed black line shows 50 $\%$ of the superconducting transition. c) The \textit{T}$_\text{c}$ values at different applied fields were used to calculate $\mu_0$\textit{H}$_\text{c2}$(0) to be 11.6 T for TaRh$_2$B$_2$ and 17.9 T for NbRh$_2$B$_2$.}
\label{Fig6}
\end{figure*}
At low temperature, the specific heat data can be described by the equation 
\begin{equation}
\frac{C_p}{T} = \gamma + \beta T^2,
\label{Eq2}
\end{equation}

\noindent where $\gamma$\textit{T} is the electronic contribution and $\beta$\textit{T}$^3$ is the phonon contribution to the specific heat.This linear relationship can be seen in the \textit{C$_p$/T} vs \textit{T}$^2$ plot in the inset of Fig.\ref{Fig5}a for TaRh$_2$B$_2$ and NbRh$_2$B$_2$. By fitting our data to the above equation (using data from the 9 T measurement taken in the normal state), the Sommerfeld parameter $\gamma$ was calculated to be 5.8(1) mJ mol$^{-1}$K$^{-2}$ for the Ta-analog and 8.8(2) mJ mol$^{-1}$K$^{-2}$ for the Nb-analog. Based on the slope of each fitted line, $\beta$ was calculated to be 0.178(5) mJ mol$^{-1}$K$^{-4}$ for TaRh$_2$B$_2$ and $\beta$ = 0.203(3) mJ mol$^{-1}$K$^{-4}$ for NbRh$_2$B$_2$. The Debye model was then used with the $\beta$ value in the equation 
\begin{equation}
\Theta_D = \left(\frac{12\pi^4}{5\beta} n R\right)^\frac{1}{3}
\label{Eq3}
\end{equation}
to calculate the Debye temperature $\Theta_D$, where n = 5 (TaRh$_2$B$_2$ or NbRh$_2$B$_2$) and \textit{R} is the gas constant 8.314 J mol$^{-1}$ K$^{-1}$. The Debye temperatures were calculated to be $\Theta_D$ = 379 K and $\Theta_D$ = 362 K for the Ta- and Nb- variant, respectively. With $\Theta_D$ and \textit{T}$_\text{c}$, the electron-phonon coupling constant, $\lambda_\text{ep}$, can then be calculated using the inverted McMillan\cite{McMillan} equation,
\begin{equation}
\lambda_{ep} = \frac{1.04 + \mu^*\ln\left( \frac{\Theta_D}{1.45T_c}\right)}{(1-0.62\mu^*)\ln\left( \frac{\Theta_D}{1.45 T_c}\right) - 1.04},
\label{Eq4}
\end{equation}
where $\mu^*$ = 0.13 and \textit{T}$_\text{c}$ = 5.8 K and 7.6 K, respectively. The superconducting parameter $\lambda_\text{ep}$ = 0.62 and 0.69 for TaRh$_2$B$_2$ and NbRh$_2$B$_2$, suggesting that both materials are weak-moderate coupling superconductors. Using $\lambda_\text{ep}$, $\gamma$, and the Boltzmann constant \textit{k}$_B$, the density of electronic states at the Fermi energy \textit{N}(\textit{E}$_\text{F}$) can be calculated from the equation
\begin{equation}
N(E_F) = \frac{3\gamma}{\pi^2 k_B^2(1 + \lambda_{ep})}.
\label{Eq5}
\end{equation}
\textit{N}(\textit{E}$_\text{F}$) was estimated to be 2.46 states eV$^{-1}$ per formula unit of TaRh$_2$B$_2$ and 3.74 states eV$^{-1}$ per formula unit of NbRh$_2$B$_2$, respectively. In addition, the normalized specific heat jump, \textit{$\Delta$C/$\gamma$T$_c$}, was found to be 1.56 and 1.60 for the Ta- and Nb-analogs, both of which are larger than the expected value of 1.43, which suggests that both materials are moderately-coupled superconductors.

Fig.\ref{Fig5}b (main panel) shows low temperature (below \textit{T}$_\text{c}$) electronic specific heat, \textit{C}$_\text{el.}$ vs. \textit{T} for NbRh$_2$B$_2$ in zero applied field and fitted to the equation
\begin{equation}
C_{el} = \gamma_0T + a* e^{-\frac{\Delta_0}{k_B T}},
\label{Eq6}
\end{equation}
where \textit{k}$_\text{B}$ is the Boltzmann constant, $\gamma_0$\textit{T} is the electronic contribution to the specific heat coming from the sample impurities, and $\Delta_0$ is the magnitude of the superconducting gap. \textit{C}$_\text{el.}$ - $\gamma_0$\textit{T} plotted as a function of the normalized \textit{T}$_\text{c}$/\textit{T} for both TaRh$_2$B$_2$ and NbRh$_2$B$_2$ is shown in the inset of Fig.\ref{Fig5}b. For a weak coupling superconductor, 
\begin{equation}
2\Delta_0 = 3.5 k_B T.
\label{Eq7}
\end{equation}
For TaRh$_2$B$_2$, the obtained value $\Delta_0$ = 0.98 meV (for which 2$\Delta_0$ = 1.96 meV) is larger than the theoretical value of 2$\Delta_0$ = 1.75 meV (for \textit{T}$_\text{c}$ = 5.8 K). Likewise for NbRh$_2$B$_2$, the calculated value of $\Delta_0$ = 1.4 meV (2$\Delta_0$ = 2.8 meV) for \textit{T}$_\text{c}$ = 7.6 K is larger than the expected value of 2$\Delta_0$ = 2.29 meV, which is consistent with what was proposed previously that TaRh$_2$B$_2$ and NbRh$_2$B$_2$ are moderately-coupled superconductors.

\textbf{Resistivity}
The temperature dependent electrical resistivity measured from 300 K - 1.7 K for TaRh$_2$B$_2$ is shown in Fig.\ref{Fig6}a (main panel) and for NbRh$_2$B$_2$ in Fig.\ref{Fig6}b (main panel). The resistivity drops to zero at 5.8 K for the Ta-variant and at 7.7 K for the Nb-variant. These new materials are both poor metals and the resistivity is essentially temperature independent for the temperature range above \textit{T}$_\text{c}$. The dependence of \textit{T}$_\text{c}$, taken as 50 $\%$ of the resistivity transition (black dashed line), on the applied magnetic field is shown in the inset of Fig.\ref{Fig6}a for TaRh$_2$B$_2$ and Fig.\ref{Fig6}b for NbRh$_2$B$_2$ where the field is increased from 0 T - 12 T in 0.5 T increments for the Ta-variant and from 0 T - 14 T for the Nb-variant. Even with a 11 T applied magnetic field, the critical temperature was only suppressed to $\sim$1.9 K for the Ta-variant and to $\sim$3 K for the Nb-variant in a 14 T applied magnetic field. Fig.\ref{Fig6}c shows the upper critical fields $\mu_0$\textit{H}$_\text{c2}$ plotted as a function of the estimated \textit{T}$_\text{c}$ values and fitted to a line close to \textit{T}$_\text{c}$ for TaRh$_2$B$_2$ and NbRh$_2$B$_2$. The resulting slope (d$\mu_0$\textit{H}$_\text{c2}$/dT) is -2.9 T/K for TaRh$_2$B$_2$ and -3.4 T/K for NbRh$_2$B$_2$. The 0 K upper critical field, $\mu_0$\textit{H}$_\text{c2}$(0), can then be estimated using the Werthamer-Helfand-Hohenberg (WHH) equation,\cite{cleanlimit}
\begin{equation}
\mu_0H_{c2}(0) = -A T_c \frac{d\mu_0H_{c2}}{dT}\bigg|_{T=T_c},
\label{Eq8}
\end{equation}
where \textit{A} is 0.693 for the dirty limit (which will be discussed later). Using \textit{T}$_\text{c}$ = 5.8 K for the Ta-analog and \textit{T}$_\text{c}$ = 7.6 K for the Nb-analog, the dirty limit $\mu_0$\textit{H}$_\text{c2}$(0) values were calculated to be 11.6 T and 17.9 T, respectively. Both of which are high and even exceed the Pauli limit ($\mu_0$\textit{H}$^\text{Pauli}$ = 1.85*\textit{T}$_\text{c}$), where $\mu_0$\textit{H}$^\text{Pauli}$ = 10.7 T for TaRh$_2$B$_2$ and $\mu_0$\textit{H}$^\text{Pauli}$ = 14.1 T for NbRh$_2$B$_2$. Even with a 14 T applied magnetic field, (which is essentially equal to the Pauli limit for NbRh$_2$B$_2$) the \textit{T}$_\text{c}$ was only suppressed to 3 K, confirming a $\mu_0$\textit{H}$_\text{c2}$(0) value that exceeds the Pauli limit. Likewise, for TaRh$_2$B$_2$, the last measurable \textit{T}$_\text{c}$ value was 1.9 K at 11 T, which is already above the Pauli limit for this material.  A higher than expected $\mu_0$\textit{H}$_\text{c2}$(0) value has also been seen in non-centrosymmetric superconductor CePt$_3$Si\cite{CePt3Si_1, CePt3Si_2}, as discussed previously, which is a good indication that the superconductors reported here are anomalous. The determined $\mu_0$\textit{H}$_\text{c2}$(0) and \textit{H}$_\text{c1}$(0) values can be used to calculate several other superconducting parameters. Using the equation 
\begin{equation}
H_{c2}(0) = \frac{\Phi_0}{2\pi\xi_{GL}^2},
\label{Eq9}
\end{equation}
\begin{table}
\caption{Observed superconductivity parameters of TaRh$_2$B$_2$ and NbRh$_2$B$_2$.}
\begin{tabular}{cccc} 

\hline 
Parameter & Units & TaRh$_2$B$_2$ & NbRh$_2$B$_2$\\

\hline 
\textit{T}$_\text{c}$							& K		&  5.8 	& 7.6 		\\
$\mu_0$\textit{H}$_\text{c1}$(0) 			& mT	&  9.6	& 13.5		\\  
$\mu_0$\textit{H}$_\text{c2}$(0) 			& T		&  11.6	& 17.9			\\ 
$\mu_0$\textit{H}$_\text{c}$(0) 				& mT	&  169	& 248		\\ 
$\xi_\text{GL}$ 				& $\textsc{\AA}$	&  53	& 43 \\ 
$\lambda_\text{GL}$ 			& $\textsc{\AA}$	&  2586		& 2190 \\
$\kappa_\text{GL}$ 			& -	&  48		& 51 \\
$\gamma$ 			& mJ mol$^{-1}$K$^{-2}$	&  5.8   & 8.8\\
\textit{$\Delta$C/$\gamma$T$_c$} 			& -	&  1.56 		& 1.60\\
$\mu_0$\textit{H}$^\text{Pauli}$ 			& T	&  10.7		& 14.1 \\ 
$\lambda_\text{ep}$ 			& -	&  0.62			& 0.69 \\
\textit{N}(\textit{E}$_\text{F}$)			& states eV$^{-1}$ per f.u.	&  2.46		& 3.74 \\
$\Theta_D$				& K				& 379 			&    362 \\
$\Delta_0$						&meV			& 0.98			& 1.4\\

\hline
\hline 
\end{tabular}
\label{Table3} 
\end{table}
where $\Phi_0$ is the quantum flux \textit{h}/2\textit{e}, the Ginzburg-Landau coherence length $\xi_\text{GL}$ was calculated to be 53 $\textsc{\AA}$ for TaRh$_2$B$_2$ and 43 $\textsc{\AA}$ for NbRh$_2$B$_2$. The dirty limit (discussed previously) was used to determine $\mu_0$\textit{H}$_\text{c2}$(0) since the obtained ratio of the coherence length $\xi_\text{GL}$ and the mean free path (\textit{l}) was greater than 1. The mean free path was determined using the following equation (derived in Ref.\cite{OsB2}),
\begin{equation}
l = 2.372\times10^{-14}\frac{\left(\frac{m^*}{m_e}\right)^2V_M^2}{N(E_F)^2\rho},
\label{Eq10}
\end{equation}
where \textit{V}$_M$ is the molar volume, $\rho$ is the resistivity, and \textit{N}(\textit{E}$_\text{F}$) is the density of states at the Fermi level. For TaRh$_2$B$_2$, \textit{V}$_M$ = 33.6 cm$^3$ mol$^{-1}$, $\rho$ = 2 m$\Omega$-cm, \textit{N}(\textit{E}$_\text{F}$) = 2.46 states eV$^{-1}$ per f.u, and assuming that m$^*$/m$_e$  = 1, we obtain \textit{l} = 22 $\textsc{\AA}$. Similarly, for NbRh$_2$B$_2$, \textit{V}$_M$ = 39.0 cm$^3$ mol$^{-1}$,  $\rho$ = 2.2 m$\Omega$-cm, and \textit{N}(\textit{E}$_\text{F}$) = 3.74 states eV$^{-1}$ per f.u, which gives \textit{l} = 12 $\textsc{\AA}$. The resulting ratios of $\xi_\text{GL}$/\textit{l} are 2.4 and 3.7 for the Ta- and Nb-variant, respectively, showing that both TaRh$_2$B$_2$ and NbRh$_2$B$_2$ are in the dirty limit. 

Using the result of $\xi_\text{GL}$ with \textit{H}$_\text{c1}$ (determined previously), the superconducting penetration depth $\lambda_\text{GL}$ = 2586 $\textsc{\AA}$ for the Ta-analog and 2190 $\textsc{\AA}$ for the Nb-analog were estimated using the lower critical field equation:
\begin{equation}
H_{c1} = \frac{\Phi_0}{4\pi\lambda_{GL}^2}\ln\frac{\lambda_{GL}}{\xi_{GL}}.
\label{Eq11}
\end{equation}
The value $\kappa_\text{GL}$ = $\lambda_\text{GL}$/$\xi_\text{GL}$ was calculated to be $\kappa_\text{GL}$ = 48 for TaRh$_2$B$_2$ and $\kappa_\text{GL}$ = 51 for NbRh$_2$B$_2$, confirming type-II superconducting behavior in both new materials. Combining the results of \textit{H}$_\text{c1}$, \textit{H}$_\text{c2}$, and $\kappa_\text{GL}$, the thermodynamic critical field can be estimated from the equation, 
\begin{equation}
H_{c1}H_{c2} = H_c^2 \ln\kappa_{GL}.
\label{Eq12}
\end{equation}

\noindent This calculation yields $\mu_0$\textit{H}$_\text{c}$ = 169 mT for the Ta-variant and 248 mT for the Nb variant. Table \ref{Table3} gives a summary of all the calculated superconducting parameters for the chiral, non-centrosymmetric superconductors TaRh$_2$B$_2$ and NbRh$_2$B$_2$. 

\textbf{Electronic Structure}
\begin{figure}[b]
\includegraphics[scale = 0.27]{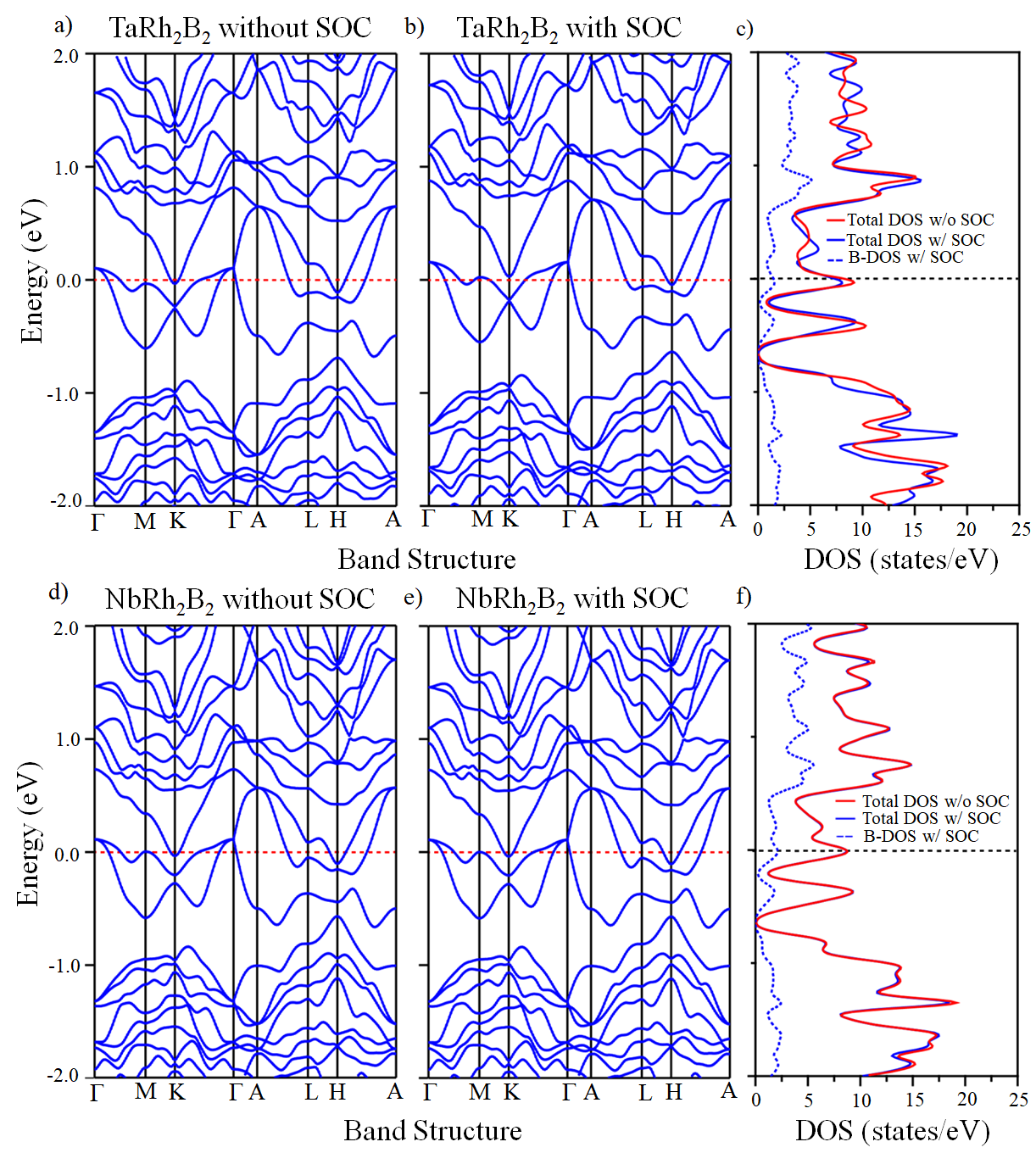}
\caption{Calculated band structure for a-b) TaRh$_2$B$_2$ and d-e) NbRh$_2$B$_2$ both with (middle) and without (left) spin-orbit coupling plotted in the energy range from -2.0 to 2.0 eV. The total density of states (DOS) calculated using the Wien2K with LDA-type pseudopotentials with spin-orbit coupling included (blue line) and without SOC (red line) for c) TaRh$_2$B$_2$ and f) NbRh$_2$B$_2$.}
\label{Fig7}
\end{figure}

To gain an intrinsic insight into the relationship between the superconductivity observed and the electronic states of TaRh$_2$B$_2$ and NbRh$_2$B$_2$, we investigated the electronic density of states (DOS) and band structure (BS) for both superconducting materials without and with spin-orbit coupling (SOC) (Fig.\ref{Fig7}). The total and partial density of states for TaRh$_2$B$_2$ and NbRh$_2$B$_2$ are illustrated in Fig.\ref{Fig7}c and Fig.\ref{Fig7}f. The DOS in the energy below -2.0 eV mainly consists of Ta/Nb and Rh \textit{s}- and \textit{d}- orbitals. The DOS in the energy range from -2.0 eV to +2.0 eV contains mixed Ta/Nb and Rh \textit{s}- and \textit{d}- orbitals in addition to \textit{s}- and \textit{p}- orbitals from B, in particular, around the Fermi level. A sharp peak in the DOS is often taken to be an indication of a nearby structural, electronic, or magnetic instability such as superconductivity. To investigate further, we calculated the band structure both with/without spin-orbit coupling for TaRh$_2$B$_2$ (Fig.\ref{Fig7}a-b) and for NbRh$_2$B$_2$ (Fig.\ref{Fig7}d-e). The broad peak in the DOS at E$_F$ is due to the presence of saddle points in the electronic structure at the \textit{M} and \textit{L} points in the Brillouin zone. Saddle points near E$_F$ are often proposed to be important for yielding superconductivity\cite{vanHove_cuprate} and may also be significant in the current materials.

\section{Conclusions}
In summary, we report two new non-centrosymmetric superconductors, TaRh$_2$B$_2$ and NbRh$_2$B$_2$, which both have a new crystal structure belonging to the chiral space group \textit{P} 3$_1$. Temperature-dependent electrical resistivity, magnetic susceptibility, and specific heat data confirmed bulk superconductivity with \textit{T}$_\text{c}$ = 5.8 K for the Ta-analog and \textit{T}$_\text{c}$ = 7.6 K for the Nb-analog. The derived superconducting parameters show that TaRh$_2$B$_2$ and NbRh$_2$B$_2$ are basically type-II BCS moderately-coupled superconductors. However, their behavior under applied magnetic fields shows that $\mu_0$\textit{H}$_\text{c2}$(0) exceeds the Pauli limit for both superconducting materials, suggesting a possible triplet superconducting state, but further study would be required to confirm this. Future work on these materials to determine the effects of their chiral non-centrosymmetric symmetry on their normal and superconducting properties, and on their use in advanced devices, particularly those with chirality junctions, will be of future interest.

\section*{Acknowledgements}
The materials synthesis was supported by the Department of Energy, Division of Basic Energy Sciences, Grant No. DE-FG02-98ER45706, and the property characterization was supported by the Gordon and Betty Moore Foundation EPiQS initiative, Grant No. GBMF-4412.  The work at LSU was supported by the Board of Regents Research Competitiveness Subprogram (RCS) under Contract No. LEQSF(2017-20)-RD-A-08 and LSU-startup funding. The research in Poland was supported by the National Science Centre, Grant No. UMO-2016/22/M/ST5/00435. N. P. O. acknowledges the support of the Gordon and Betty Moore Foundations EPiQS Initiative through Grant No. GBMF4539. J. J. L. was supported by funds from the National Science Foundation, NSF MRSEC Grant No. DMR 1420541. 

\section*{Author Correspondence}
\begin{flushleft}
$^*$E.M.C (carnicom@princeton.edu)\\
$^*$R.J.C. (rcava@exchange.princeton.edu)
\end{flushleft}

\bibliographystyle{apsrev4-1}
\bibliography{MyBib}

\end{document}